\documentstyle[12pt]{article}
\def \ms {{\overline{\mbox{MS}}}}

\newcommand{\titul}[1] {\begin{center}{\large\bf #1 } \end{center}\vskip 1.cm}
\newcommand{\autor}[1] {\begin {center} {\large \lineskip .5em #1 }
                        \end   {center} }
\newcommand{\lugar}[1] {\begin{center} {\it #1} \end{center}}
\newcommand{\abstr}[1] {{\begin{center} \vskip .5cm {\bf Abstract
                        \vspace{0pt}} \end{center}}\begin{quote} #1
                        \end{quote}}
\newcommand{\z}{&&\hspace*{-1cm}}
\newcommand{\ep}{\varepsilon}
\begin{document}
\begin{titlepage}
.
\begin{flushright} {\bf US-FT/13-98 \\ July 4th, 1998} \end{flushright}

\vskip 3.cm
\titul{ Small $x$ behaviour of parton distributions\\
with soft initial conditions
}

\autor{A.V. Kotikov\footnote{E-mail:KOTIKOV@SUNSE.JINR.DUBNA.SU}}
\lugar{Particle Physics Laboratory\\
Joint Institute for Nuclear Research\\
141980 Dubna, Russia}
\autor{G. Parente\footnote{E-mail:GONZALO@GAES.USC.ES}}
\lugar{Departamento de F\'\i sica de Part\'\i culas\\
Universidade de Santiago de Compostela\\
15706 Santiago de Compostela, Spain}
\abstr{
We present an analytical parametrization of the QCD
description of the small $x$ behaviour
of parton distribution functions in the leading twist approximation
of the Wilson operator product expansion, in the case
of soft initial conditions.
The results are in very good agreement with deep inelastic scattering
experimental data from HERA.

PACS number(s): 13.60.Hb, 12.38.Bx, 13.15.Dk

}
\end{titlepage}
\newpage

\pagestyle{plain}

\section{Introduction} \indent

The measurements of the deep-inelastic scattering (DIS)
structure function (SF) $F_2$ in HERA \cite{H1,ZEUS,ZEUSB}
have permitted the access to
a very interesting kinematical range for testing the theoretical
ideas on the behavior of quarks and gluons carrying a very low fraction
of momentum of the proton, the so-called small $x$ region.
In this limit one expects that
nonperturbative effects may give essential contributions. However, the
resonable agreement between HERA data and the NLO approximation of
perturbative
QCD has been observed for $Q^2 > 1 $GeV$^2$ (see the recent review
in \cite{CoDeRo}) and, thus,
perturbative QCD could describe the
evolution of structure functions up to very low $Q^2$ values,
traditionally explained by soft processes.
It is of fundamental importance to find out the kinematical region where
the well-established perturbative QCD formalism
can be safely applied at small $x$.

The standard program to study the small $x$ behavior of
quarks and gluons
is carried out by comparison of data
with the numerical solution of the
Dokshitzer-Gribov-Lipatov-Altarelli-Parisi (DGLAP)
equations 
\cite{DGLAP1, DGLAP} \footnote{ At small $x$ there is another approach
based on the Balitsky-Fadin-Kuraev-Lipatov (BFKL) equation \cite{BFKL}, whose
application is out of the scope of this work.} by
fitting the parameters of the
$x$ profile of partons at some initial $Q_0^2$ and
the QCD energy scale $\Lambda$ \cite{fits}-\cite{GRV}.
However, if one is interested in analyzing exclusively the
small $x$ region, there is the alternative of doing a simpler analysis
by using some of the existing analytical solutions of DGLAP 
in the small $x$
limit \cite{BF1}-\cite{Munich2}.
This was done so in Ref. \cite{BF1,BF2}
where it was pointed out that the HERA small $x$ data can be
interpreted in 
terms of the so called doubled asymptotic scaling phenomenon
related to the asymptotic 
behaviour of the DGLAP evolution 
discovered  in \cite{DGLAP1,Rujula} many years ago. 

On the other hand, various groups have been able to fit
the available data (essentially at large $Q^2$)
using a hard input at small $x$: 
$x^{-\lambda},~\lambda >0$.
In some sense, it is not very surprising, because the 
modern HERA data cannot distinguish yet between the behavior
of a steep input parton distribution and the
quite steep dynamical evolution from a soft initial condition. 
Moreover, for the full set  of
anomalous dimensions (AD) obtained at $x \to 0$ in Ref. \cite{GSFM}
based on BFKL, the results weakly 
depend on the form of the initial condition (see \cite{Catani}),
preserving the hard ones and changing the soft ones.
In the case of working with fixed order AD, the initial conditions are 
important when the data are considered in a wide range of $Q^2$
and  it is necessary to choose the form of the PD asymptotics
at some $Q^2_0$.
In this work we use
a soft initial condition
in agreement with the experimental situation: low-$Q^2$ data
\cite{NMC,lowQ2} are well described at $Q^2 \leq 0.4$ GeV$^2$ by
Regge theory 
with Pomeron intercept $\ep_P \equiv \lambda +1 =1.08$,
closed to the standard ($\ep_P =1$) one. Moreover, HERA data \cite{lowQ2}
with $Q^2 > 1$ GeV$^2$ are in good agreement with GRV
predictions \cite{GRV1,GRV} which support our aim to develop an
analytical form for the parton densities at small $x$ because,
at least conceptually, it
is very closed to the GRV approach.

Thus, the purpose of this article is to obtain the small $x$ asymptotic
form of parton distributions (PD)
in the framework of the DGLAP equation starting at some $Q^2_0$ with
the flat function:
 \begin{eqnarray}
f_a (Q^2_0) ~=~
A_a ~~~~(a=q,g), \label{1}
 \end{eqnarray}
where $f_a$ are the parton distributions multiplied by $x$
and $A_a$ are unknown parameters that have to be determined from data.
Through this work at small $x$ we neglect
the non-singlet quark component.

The article is organized as follows. Sections 2 and 3 contain LO analyses:
in Sect. 2 we consider the simple but important case without quarks,
while in Sect. 3 the quarks distributions are taken into account.
The NLO analysis, which is the main 
result of this article, is performed in Sect. 4. In Sect. 5 we present
the fits to experimental data and some discussions of the obtained results.
In the Appendix we illustrate
the method \cite{method} of 
replacing at small$~x$ the convolution of two functions
by a simple product. The method is used in the present work for the
correct incorporation of the non-singular part of parton distributions
to our formulae.

\section{Leading order without quarks} \indent

First of all, we consider the leading order (LO) approximation without quarks
as a pedagogical example of the more cumbersome calculations below.
This case is at the same time very simple and very closed
to the real situation, because gluons give the basic contribution at small
$x$.

At the momentum space, the solution of the DGLAP equation in this case has the form

 \begin{eqnarray}
 M_g(n,Q^2)~=~M_g(n,Q^2_0)e^{-d_{gg}(n)  s},
 \label{2}
 \end{eqnarray}
where $M_g(n,Q^2)$ are the moments of the gluon distribution,
 $$ s=ln\left(\frac{\alpha(Q^2_0)}{\alpha(Q^2)}\right)
~~~\mbox{ and }~~~d_{gg}=\frac{\gamma^{(0)}_{gg}(n)}{2\beta_0}$$

The terms
$\gamma^{(0)}_{gg}(n)$ and $\beta_0$ are respectively the LO
coefficients of the gluon-gluon AD and the QCD $\beta $-function.
Through this work we use the short notation 
$\alpha(Q^2) = \alpha_s(Q^2) /(4 \pi)$.

At LO, $s$ can be written in terms of the QCD scale $\Lambda$ as:

 \begin{eqnarray}
s_{LO}=ln\left(\frac{ln(Q^2/\Lambda^2_{LO})}
{ln(Q^2_0/\Lambda^2_{LO})}\right)
 \label{2.5}
 \end{eqnarray}

For any perturbatively calculable variable $K(n)$,
it is very convenient to separate the singular part when 
$n \to 1$ (denoted by ``$\widehat{K}$'') and the regular part
(marked as ``$\overline K$'').
Then,  Eq. (\ref{2}) can be represented by the form

 \begin{eqnarray}
 M_g(n,Q^2)~=~M_g(n,Q^2_0)e^{-\hat d_{gg} s_{LO}/(n-1)}
e^{-\overline d_{gg}(n) s_{LO}},
 \label{3}
 \end{eqnarray}
with $\hat \gamma_{gg} = -8 C_A $ and $C_A=N$ for $SU(N)$ group.

Because we are only interested in
the small $x$ behaviour and the initial conditions are given
by the soft ($x$-independent) functions in Eq. (\ref{1}),
we use permanently the variable
$z=x/x_0$ with values $0<z<1$ with some 
arbitrary $x_0 \leq 1$
\footnote{
The correct incorporation of the regular part of the parton distributions
can be done only for small values of $z$ (see Eq. (\ref{5})) . Thus, $x_0$
should be restricted to the range
$0.1 \leq x_0 \leq 1$ in order to keep the correctness
of our formulae at $x \leq 10^{-2}$.}, i.e.
$$ M_a(n,Q^2)~=~ \int^1_0 dz z^{n-2} f_a(z,Q^2)$$.

Finally, if one takes the soft boundary conditions given by
Eq. (\ref{1}),  the coefficient in Eq. (\ref{3}) becomes
 \begin{eqnarray}
 M_a(n,Q^2_0)~=~\frac{A_a}{n-1}
 \label{3.1}
 \end{eqnarray}

\subsection{Classical double-logarithmic case } \indent

As a first step, we consider the classical double-logarithmic case which
corresponds to use only the AD singular part, i.e. 
$\overline d_{gg}(n) =0$ in Eq. (3).

Then, expanding the second exponential in the r.h.s. of Eq. (3) 
$$ M_g^{cdl}(n,Q^2)~=~A_g \sum^{\infty}_{k=0} \frac{1}{k!}
\frac{{(-\hat d_{gg} s_{LO})}^k}{(n-1)^{k+1}} $$
and 
using the Mellin transformation for $\Bigl(ln(1/z)\Bigr)^k$:

$$  \int^1_0 dz z^{n-2} {(ln(1/z))}^k ~=~ \frac{k!}{(n-1)^{k+1}}$$
we
immediately obtain the well known double-logarithmic behavior 
\cite{DGLAP1,Rujula}:

 \begin{eqnarray}
f_g^{cdl}(z,Q^2)~=~A_g \sum^{\infty}_{k=0} \frac{1}{(k!)^2}
{(-\hat d_{gg}s_{LO})}^k {(ln(1/z))}^k ~=~ A_g I_0(\sigma_{LO}),
 \label{4}
 \end{eqnarray}
where
$I_0(\sigma_{LO})$ is the modified Bessel function with 
argument\footnote{Hereafter
we follow the popular Ball-Forte variables 
$\sigma_{LO}$ and $\rho_{LO}$. Below, they are generalized to be used beyond
the LO approximation ($\sigma $ and $\rho $).}
 $\sigma_{LO} =2\sqrt{\hat d_{gg} s_{LO} ln(z)}$.

\subsection{The more general case } \indent

For a regular kernel $\tilde K(z)$ (see \cite{method} and Appedix),
 having Mellin transform

$$ K(n)~=~ \int^1_0 dz z^{n-2} \tilde K(z)$$
and the PD $f_a(z)$ in the form 
$I_{\nu}(\sqrt{\hat d ln(1/z)})$
we
have the following equation

 \begin{eqnarray}
\tilde K(z) \otimes f_a(z)~=~ K(1)f_a(z) + 
O\Biggl(\sqrt{\frac{\hat d}{ln(1/z)}}
\Biggr)
 \label{5}
 \end{eqnarray}

The Eq. (\ref{5}) has been obtained for the nonsingular (at 
$n \to 1$)
functions  $K(n)$ and Regge-like PD many years ago by the Madrid group 
\cite{LoYn}.
It has been expanded for arbitrary 
kernel $\tilde K(z)$ in \cite{method} and it was already used in \cite{KoPa}
and \cite{KOPAFL}
to extract the gluon distribution  and the longitudinal structure 
function $F_L$ from $F_2$ and $dF_2/dln(Q^2)$ data. For 
arbitrary $\tilde K(z)$ the ability
of the method \cite{method} has been checked numerically in Refs. \cite{KoPa}
and \cite{KOPAFL} using MRS sets of PD.

With the help of Eq. (\ref{5}) 
one can find the general solution for the
LO gluon density without the influence of quarks

 \begin{eqnarray}
f_g(z,Q^2)~=~ A_g I_0(\sigma_{LO}) e^{-\overline d_{gg}(1) s_{LO}} 
~+~O(\rho_{LO}),
 \label{6}
 \end{eqnarray}
where
$$\rho_{LO} = \sqrt{\frac{\hat d_{gg} s_{LO}}{ln(z)}}
= \frac{\sigma_{LO}}{2ln(1/z)}~,~~~ 
\overline \gamma^{(0)}_{gg}(1) = 
22 +\frac{4}{3}f ~~~ \mbox{ and } ~~~
\overline d_{gg}(1) = 1+\frac{4f}{3\beta_0}$$ with $f$ as the number of active 
quarks.

\section{Leading order (complete)} \indent

At the momentum space, the solution of the DGLAP equation at LO
has the form

 \begin{eqnarray}
 M_a(n,Q^2) &=& M_a^+(n,Q^2) + M_a^-(n,Q^2) ~\mbox{ and }~ \nonumber \\
 M_a^{\pm}(n,Q^2) &=& M_a^{\pm}(n,Q^2_0)e^{-d_{\pm}(n) s} = M_a^{\pm}
e^{-\hat d_{\pm} s/(n-1)}e^{-\overline d_{\pm}(n) s},
 \label{6.1}
 \end{eqnarray}
where\footnote{We use a non-standard definition
(see \cite{YF93})
of the projectors $\ep_{ab}^{\pm}(n)$, which is very convenient
beyond LO (see Eq. (\ref{8.22})). The connection with the more usual
definition $\alpha$, $\tilde \alpha$ and $\ep$ in ref.
\cite{Reya,Buras} is given by: $\ep_{qq}^{-}(n)= \alpha (n)$,
$\ep_{qg}^{-}(n)= \tilde \alpha (n)$ and $\ep_{gq}^{-}(n)= \ep (n)$ }
 \begin{eqnarray}
 M_a^{\pm}(n,Q^2) &=&  \ep_{ab}^{\pm}(n) M_b(n,Q^2),~~~~
d_{ab}=\frac{\gamma^{(0)}_{ab}(n)}{2\beta_0}, \nonumber \\
d_{\pm}(n)&=& \frac{1}{2}
\Biggl[
\Bigl(d_{gg}(n)+d_{qq}(n) \Bigr)\pm  \Bigl( d_{gg}(n)- d_{qq}(n) \Bigr)
\sqrt{ 1
  +\frac{4d_{qg}(n)d_{gq}(n)}{( d_{gg}(n)- d_{qq}(n))^2}}
\Biggr] \nonumber \\
 \ep_{qq}^{\pm}(n)&=&  \ep_{gg}^{\mp}(n) = \frac{1}{2}
\biggl( 1+ \frac{ d_{qq}(n)- d_{gg}(n)}{d_{\pm}(n)- d_{\mp}(n)} \biggr),~ 
  \ep_{ab}^{\pm}(n) = \frac{ d_{ab}(n)}{d_{\pm}(n)- d_{\mp}(n)}
(a \neq b)
 \label{6.2}
 \end{eqnarray}

As the singular (when $n \to 1$) part of the $+$ component of the
anomalous dimension is $\hat d_{+}=\hat d_{gg} = -4 C_A/\beta_0$ 
while the $-$ component does not exist ($\hat d_{-}=0$),
we consider below both cases separately.\\

\subsection{The ``$+$'' component } \indent

The analysis of the ``$+$''
component is practically identical to the case studied in section 2. 
The only difference lies in the appearance of new terms
$ \ep_{ab}^{+}(n) $. 
If they are expanded in the vicinity of $n=1$ in the form
$ \ep_{ab}^{+}(n) =  \overline \ep_{ab}^{+} + (n-1) \tilde \ep_{ab}^{+} $,
then for the terms  $\overline \ep_{ab}^{+}$ multiplying 
$M_b(n,Q^2)$, we have the same results as in previous section:
$$
\overline \ep_{ab}^{+} M_b(n,Q^2) ~
\stackrel{ {\cal{M}}^{-1} }{\longrightarrow}
~ \overline \ep_{ab}^{+} 
A_b I_0(\sigma_{LO}) e^{-\overline d_{+}(1) s_{LO}} ~+~ O(\rho_{LO}),
$$
where
the symbol $\stackrel{{\cal{M}}^{-1} }{\longrightarrow}$ denotes the
inverse Mellin transformation.
The values of $\sigma$ and $\rho$ coincide with those defined 
in the previous section because $\hat d_{+} = \hat d_{gg}$.

The terms $\tilde \ep_{ab}^{+} $ 
that come with the additional factor $(n-1)$ in front,
lead to the following results
 \begin{eqnarray} \z
(n-1) \tilde \ep_{ab}^{+} \frac{A_b}{(n-1)} 
e^{-\hat d_{+} s_{LO}/(n-1)} =
\tilde \ep_{ab}^{+} A_b \sum^{\infty}_{k=0} \frac{1}{k!}
\frac{{(-\hat d_{+} s_{LO})}^k}{(n-1)^{k}} \nonumber \\
\z 
\stackrel{{\cal{M}}^{-1} }{\longrightarrow}
~  \tilde \ep_{ab}^{+} A_b
\sum^{\infty}_{k=0} \frac{1}{k!} \frac{1}{(k-1)!}
{(-\hat d_{+}s_{LO})}^k {(ln(1/z))}^{k-1} ~=~ 
\tilde \ep_{ab}^{+} A_b
\rho_{LO} I_1(\sigma_{LO}), \nonumber 
 \end{eqnarray}
i.e.
the additional factor $(n-1)$ in momentum space leads to replacing the Bessel 
function
$ I_0(\sigma_{LO})$ by $\rho_{LO} I_1(\sigma_{LO})$ in $z$-space.

Thus, we obtain that the term 
$ \ep_{ab}^{+}(n) M_b(n,Q^2)$ leads to the following contribution in $z$ space:
 \begin{eqnarray}
\biggl(
 \overline \ep_{ab}^{+} I_0(\sigma_{LO}) + 
\tilde \ep_{ab}^{+} 
\rho_{LO} I_1(\sigma_{LO}) \biggr)
A_b  e^{-\overline d_{+}(1) s_{LO}} ~+~ O(\rho_{LO})
 \label{7}
 \end{eqnarray}

Because the Bessel function $I_{\nu}(\sigma)$ has the $\nu$-independent 
asymptotic behavior
$e^{(\sigma)}/\sqrt{\sigma}$ at $\sigma \to \infty$ (i.e. $z \to 0$), 
the second term in
Eq. (\ref{7}) is $O(\rho)$ and must be kept only when
$ \overline \ep_{ab}^{+} =0$. This is the case for the quark distribution
at the LO approximation.

Using the concrete AD values, one has
 \begin{eqnarray}
f_g^+(z,Q^2)&=& \biggl(A_g + \frac{4}{9} A_q \biggl)
I_0(\sigma_{LO}) e^{-\overline d_{+}(1) s_{LO}} ~+~O(\rho_{LO}) 
~~\mbox { and } \nonumber \\
f_q^+(z,Q^2)&=& \frac{f}{9}\biggl(A_g + \frac{4}{9} A_q \biggl) \rho_{LO}
I_1(\sigma_{LO}) e^{-\overline d_{+}(1) s_{LO}} ~+~O(\rho_{LO}) \label{8.0}
 \end{eqnarray}
where
$\overline d_{+}(1) = 1+20f/(27\beta_0)$.

\subsection{the ``$-$'' component } \indent

In this case the anomalous dimension is regular and using 
Eq. (\ref{5})\footnote{In the 
Regge-like case (see \cite{method} and Appendix) 
$ f_a(x) \sim x^{-\lambda} \sim z^{-\lambda}$ the
Eq. (\ref{5}) has $K(1+\lambda)$ in its r.h.s. with
the accuracy $O(z)$ for an arbitrary $\lambda$, 
including the case $\lambda=0$.} one has
$$
\ep_{ab}^{-}(n) A_b e^{-d_{-}(n) s}~
\stackrel{{\cal{M}}^{-1} }{\longrightarrow}
 ~ \overline \ep_{ab}^{-}(1)
A_b  e^{-d_{-}(1) s_{LO}} ~+~ O(z)
$$
for PD.

Using the concrete AD values, we have
 \begin{eqnarray}
f_g^-(z,Q^2)&=& - \frac{4}{9} A_q e^{- d_{-}(1) s_{LO}} 
~+~O(z) ~\mbox { and } 
\nonumber \\
f_q^-(z,Q^2)&=&  A_q e^{- d_{-}(1) s_{LO}} ~+~O(z),
 \label{8.00}
 \end{eqnarray}
where
$\overline d_{-}(1) = 16f/(27\beta_0)$.\\

Finally we present the full small $x$ asymptotic results
for PD and $F_2$ structure function at LO of perturbation theory:

 \begin{eqnarray}
 f_a(z,Q^2) &=& f_a^+(z,Q^2) + f_a^-(z,Q^2) ~\mbox{ and }~ \nonumber \\
F_2(z,Q^2)&=& e \cdot f_q(z,Q^2)
\label{8}
\end{eqnarray}
where $f_q^+$,$f_g^+$, $f_q^-$ and $f_g^-$ are given by Eqs. (\ref{8.0}) and
(\ref{8.00})
and
$e= \sum^f_1e^2_i/f$  is the average charge square of the $f$ active quarks.

Let us now describe the main conclusions that follow Eq. (\ref{8}).

\begin{itemize}
\item Our LO results for PD coincide with the ones obtained
by Mankievitcz et al. in Ref. \cite{Munich1}
for $f=4$ active quarks.
The ``$+$'' component part coincides with the Ball and Forte LO
results \cite{BF1}.
\item 
The
``$+$'' and ``$-$'' components are presented explicitly separated.
The ``$-$'' component $\sim Const$ is negligible  at small $x$ (and 
large $Q^2$) in comparison 
with $\rho_{LO}I_1(\sigma_{LO})$ (as it has been already observed in 
\cite{YF93,BF1,Munich1}) and the LO quark 
distribution is ``driven'' by the influence of the gluons:
$f_q^+(z,Q^2) \approx (f/9) \rho
f_g^+(z,Q^2)$ (see also \cite{YF93, BF1, KOTILOWX, Munich1}). 
However, at intermediate $Q^2$, the
``$-$'' component is essential (as it was discussed in 
\cite{KOTILOWX,Munich1}).
Thus, in order to give the more general result valid for a wide $Q^2$ range, 
we consider PD (see Eq. (\ref{8})) as the combinations of the
``$+$'' and ``$-$'' components, where every component evolves
{\it independently}.
\item The separation of the singular and regular parts of the AD
performed above
leads to the possibility of avoiding complicated methods for evaluating 
the inverse Mellin convolution or  special analyses of DGLAP equations
(see \cite{BF2} for a review on these methods). In our case,
we use the exact  solution to get the moments of the PD.
The simple form of the singular part of
this exact solution is easily transformed to the $z$-space.
The non-singular part is added by the method of replacing 
Mellin convolution by usual product \cite{method}.
In this case the non-singular part in the $z$-space
is equal to the corresponding contribution for the first moment $n=1$. 
\end{itemize}

In the following we resume the steps we have followed to reach
the small $x$ approximate solution of DGLAP shown above:

\begin{itemize}
\item Use the $n$-space exact solution.
\item Expand the perturbatively calculated parts
     (AD and coefficient functions) in the vicinity of the point $n=1$.
\item The singular part with the form 
 \begin{eqnarray}
A_a (n-1)^k e^{-\hat d s_{LO}/(n-1)} 
 \label{9.1}
 \end{eqnarray}
leads to Bessel functions in the $z$-space in the form \begin{eqnarray}
A_a {\biggl(\frac{\hat d s_{LO}}{lnz}\biggr)}^{(k+1)/2} 
I_{k+1}\biggl(2\sqrt{\hat d s_{LO} lnz}\biggr) 
 \label{9.2}
 \end{eqnarray}
\item The regular part $B(n) \exp{(-\overline d(n) s_{LO})}$ leads to the
additional coefficient (see \cite{method} and Appendix)
$$B(1) exp{(-\overline d(1) s_{LO})} + O(\sqrt{\hat d s_{LO}/lnz}) $$
 behind of the Bessel
function (\ref{9.2}) in the $z$-space. Because the accuracy is 
$O(\sqrt{\hat d s_{LO}/lnz}) $,
it is necessary to use
only the basic term of Eq. (\ref{9.2}), i.e. all terms $(n-1)^k$ in front
of $\exp{(-\hat d/(n-1))}$, with the exception of one with the
smaller $k$ value, can be neglected.
\item If the singular part at $n \to 1$ is absent, i.e. $\hat d =0$ in
(\ref{9.1}), the result in the $z$-space is determined by 
$B(1) exp{(-\overline d(1) s_{LO})}$ with accuracy $O(z) $.
\end{itemize}

We would like to stress that the applicability of the above recipe
is not limited by the order in perturbation theory but by the form
of the singular part of the anomalous dimensions.
At the first two orders of perturbation theory the singular part is
proportional to $\sim (n-1)^{-1}$
but this behaviour does not remain at higher orders.
The most singular terms
have been
calculated in \cite{GSFM, CCH}. For example, the gluon-gluon AD has the form

 \begin{eqnarray}
\gamma_{gg}(n,\alpha) = \gamma (n,\alpha) +
O\biggl(\alpha {\biggl(\frac{\alpha }{n-1}\biggr)}^k \biggr) 
 \label{9.d}
 \end{eqnarray}
where
the terms $\sim O\Bigl(\alpha (\alpha /(n-1))^k \Bigr) $
have been evaluated \cite{FL} very recently.

The BFKL anomalous dimension $\gamma (n,\alpha)$ is obtained
by solving the implicit equation  
$$
1 ~=~ \frac{4C_A \alpha }{n-1} \chi \Bigl( \gamma (n,\alpha) \Bigr), 
$$
where the characteristic function $\chi ( \gamma )$ has the following
expression in terms 
of the Euler $\Psi$-function:
$$
\chi ( \gamma  ) ~=~ 2 \Psi(1) - \Psi (\gamma ) - \Psi (1- \gamma )
$$

The expansion of $\gamma (n,\alpha)$ in powers of
$\overline \alpha_s = 4 C_A \alpha $  gives:

 \begin{eqnarray}
\gamma (n,\alpha)
\simeq
\frac{\overline \alpha_s }{n-1} + 
2.404 {\left(\frac{\overline \alpha_s }{n-1}\right)}^4 + 
2.074 {\left(\frac{\overline \alpha_s }{n-1}\right)}^6 +
O\biggl({\left(\frac{\overline \alpha_s }{n-1}\right)}^k \biggr)
 \label{9.d1}
 \end{eqnarray}
which explicitly shows the  of the 
the term $\sim (n-1)^{-4}$ in the fourth order
of the expansion.
Moreover, the dependence $\sim (n-1)^{-2}$  
has been found \cite{FL} when it  is considered the third order
terms proportional to $\Bigl(\alpha (\alpha /(n-1))^k \Bigr) $ 
in the r.h.s. of Eq. (\ref{9.d}).
Note that the regular part of this
 third order coefficient in the perturbative expansion of the AD
is only partially known\footnote{It is known
the n=2, 4, 6, 8 and 10 Mellin moments \cite{LRV} }
although there is the possibility to calculate it completely \cite{Chetyrkin}. 

Thus, an extension of our recipe beyond the NLO approximation
requires the evaluation of singular terms $\sim (n-1)^{-k}~~(k>1)$
which is not easy and out of the scope of this work.
We restrict ourselves to
the first two orders in perturbation theory.

\section{Next-to-leading order} \indent

At the momentum space, the solution of the DGLAP equation
has the form\footnote{The representation (\ref{8.1}) for the moments
$M_a^{\pm}(n,Q^2)$ is slightly
different from the one in Ref. \cite{Buras} (see Eqs. (2.139)-(2.144) 
in \cite{Buras}) because we prefer to separate the ``$+$'' and ``$-$''
evolutions exponents exactly at NLO. The sum $M_a^+(n,Q^2) + M_a^-(n,Q^2)$ 
coincides with the one of \cite{Buras}.}
 \begin{eqnarray}
 M_a(n,Q^2) &=& M_a^+(n,Q^2) + M_a^-(n,Q^2) ~\mbox{ and }~ \nonumber \\
 M_a^{\pm}(n,Q^2) &=& \widetilde M_a^{\pm}(n,Q^2,Q^2_0)exp(-d_{\pm}(n)s 
-D_{\pm}(n)p) \label{8.1}\\
 &=& \widetilde M_a^{\pm}(n,Q^2,Q^2_0)
exp(-(\hat d_{\pm}s + \hat D_{\pm}p)/(n-1))exp(-\overline d_{\pm}(n)s
-\overline D_{\pm}(n)p),
\nonumber
 \end{eqnarray}
where
the new variable $p$ is $p=\alpha(Q^2_0)-\alpha(Q^2)$.

In comparison with the LO expressions, the NLO AD leads to the
following additional factors in the moments:
\begin{itemize}
\item The term proportional to $p$ contributes to the evolution part.
It can be represented by
 \begin{eqnarray}
D_{\pm}(n) ~=~ d_{\pm\pm}(n) -\frac{\beta_1}{\beta_0}d_{\pm}(n) 
 \label{8.2}
 \end{eqnarray}
and analogously for its singular and regular parts.
\item The terms proportional to $\alpha(Q^2)$ and $\alpha(Q^2_0)$ change
the normalization factor \\
$ M_a^{\pm}(n,Q^2_0) \to \widetilde M_a^{\pm}(n,Q^2,Q^2_0)$:
 \begin{eqnarray}
\widetilde M_a^{\pm}(n,Q^2,Q^2_0) = \Biggl(1-
d^a_{\pm\mp}(n) 
\alpha(Q^2) \Biggr)
M_a^{\pm}(n,Q^2_0) + 
d^a_{\mp\pm}(n) \alpha(Q^2_0) M_a^{\mp}(n,Q^2_0)
 \label{8.3}
 \end{eqnarray}
\end{itemize}

The NLO components of the r.h.s. of Eqs. (\ref{8.2}, \ref{8.3}) have the form:

 \begin{eqnarray}
d_{\pm\pm}(n)&=&  \sum_{a,b=q,g} \ep^{\pm}_{ba} d^{(1)}_{ab} \nonumber \\
d_{\pm\mp}(n)&=&  \sum_{a=q,g} \ep^{\pm}_{ag} d^{(1)}_{ga} -
\ep^{\mp}_{qq} d^{(1)}_{qq} +
\Bigl(\ep^{\pm}_{gq} - \ep^{\pm}_{gg}/\ep^{\pm}_{qg} \Bigr) d^{(1)}_{gg} 
\nonumber \\
d^q_{\pm\mp}(n)&=& \frac{d_{\pm\mp}(n)}{1+d_{\pm}(n)-d_{\mp}(n)},~
d^g_{\pm\mp}(n)~=~d^q_{\pm\mp}(n)\frac{\ep^{\mp}_{gg}}{\ep^{\mp}_{qq}},
 \label{8.22}
 \end{eqnarray}
where
$d^{(1)}_{ab}(n)=\gamma^{(1)}_{ab}(n)/2\beta_0$ and  $\gamma^{(1)}_{ab}(n)$ 
are the NLO AD.

We would like to stress that the exponent in Eq. (\ref{8.1})
contains the NLO contribution proportional to Eq. (\ref{8.2}).
We have checked this result by direct calculation
of NNLO corrections to the solution of DGLAP equation (following
the review of Buras \cite{Buras}), arriving to terms with the
form $\sim D^2_{\pm}(n)/2$
\footnote{ These corrections are also needed
to extend the NNLO analysis of structure functions
\cite{NNfits} from non-singlet to singlet behaviour.}.

The exponential representation of the term proportional
to  Eq. (\ref{8.2}) is very important
in the case of the + component because it
contains a singular part when $n \to 1$ that could make it
of the order of the LO contribution, spoiling the perturbative convergence
if one attempts to expand the NLO part of the exponent.

Following the steps given in the previous section, 
one can easily obtain the small $x$ behaviour of the PD and
$F_2$ at NLO. It has the form:
 
 \begin{eqnarray}
 f_a(z,Q^2) &=& f_a^+(z,Q^2) + f_a^-(z,Q^2) ~\mbox{ and }~ \nonumber \\
f_a^-(z,Q^2)&=& A_a^-(Q^2,Q^2_0) exp(- d_{-}(1)s-D_{-}(1)p) ~+~O(z) 
\label{9.10} \\
f_g^+(z,Q^2) &=& A_g^+(Q^2,Q^2_0)
I_0(\sigma) exp(- \overline d_{+}(1)s-\overline D_{+}(1)p)
~+~O(\rho) 
\label{9.11} \\
f_q^+(z,Q^2)&=& 
A_q^+(Q^2,Q^2_0)
\biggl[ (1 - \bar{d}_{\pm}^q(1) \alpha(Q^2)) \rho I_1(\sigma)
      + 20 \alpha(Q^2) I_0(\sigma) \biggr]
\nonumber \\
& & 
\cdot 
exp(- \overline d_{+}(1)s-\overline D_{+}(1)p)
+O(\rho) \label{9.12} \\
F_2(z,Q^2)&=& e \cdot \biggl(f_q(z,Q^2) + \frac{2}{3}f\alpha(Q^2) f_g(z,Q^2)
 \biggr)
\label{9}
\end{eqnarray}
where 
 \begin{eqnarray}
\sigma &=& 2\sqrt{(\hat d_{+}s+\hat D_{+}p)lnz} ~~, ~~~~
\rho = \sqrt{\frac{(\hat d_{+}s+\hat D_{+}p)}{lnz}}=
\frac{\sigma }{2ln(1/z)}
~~,
\label{a1}\\
A_g^+(Q^2,Q^2_0) &=& \Bigr[1-\frac{80}{81}f\alpha(Q^2) \Bigr]A_g 
+ \frac{4}{9}\Bigl[1+3(1+\frac{1}{81}f)\alpha(Q^2_0) - 
\frac{80}{81}f  \alpha(Q^2)
\Bigr] A_q ~~, 
\nonumber \\
A_g^-(Q^2,Q^2_0) &=& A_g - A_g^+(Q^2,Q^2_0)  \nonumber \\
A_q^+ &=& \frac{f}{9}\biggl(A_g + \frac{4}{9} A_q \biggl)~~, ~~~
A_q^- = A_q - 20 \alpha(Q^2_0) A_q^+
\label{a2} \end{eqnarray}

The components of the singular and regular parts of $D_{\pm}$ have the form:

 \begin{eqnarray}
\hat d_{++} &=& \frac{412}{27\beta_0}f ~~, ~~~
\hat d^q_{+-} = -20 ~~, ~~~
\hat d^g_{+-} = 0~~, \nonumber \\
\overline d_{++}(1) &=& \frac{8}{\beta_0}
\biggl( 36 \zeta_3 + 33 \zeta_2 - \frac{1643}{12} +\frac{2}{9}f 
\Bigr[ \frac{68}{9} -4 \zeta_2 - \frac{13}{243}f \Big] \biggr)~~, \nonumber \\
\overline d^q_{+-}(1) &=& 
\frac{134}{3} -12 \zeta_2 - \frac{13}{81}f 
~~, ~~~
\overline d^g_{+-}(1) = \frac{80}{81}f~~, \nonumber \\
d_{--}(1) &=& \frac{16}{9\beta_0}
\biggl( 2 \zeta_3 - 3 \zeta_2 + \frac{13}{4} + f 
\Bigr[  4 \zeta_2 - \frac{23}{18} + \frac{13}{243}f \Big] \biggr)~~,
\nonumber \\
d^q_{-+}(1) &=& 0~~, ~~~
d^g_{-+}(1) = -3 \Bigl( 1+ \frac{f}{81} \Bigr)~~.
 \label{9.3}
 \end{eqnarray}

The numerical values of these coefficients are listed in Tab. 1 for
a different number of active quarks.

The Eqs. (\ref{9.10})-(\ref{9.3}) together with the recipes in the end
of the section 3 are the main results of this article.\\

Looking carefully Eqs. (\ref{9.10})-(\ref{9}), we arrive to the following 
conclusions:
\begin{itemize}
\item Our NLO results coincide with the corresponding of Ball
and Forte in Ref. \cite{BF2} if one neglects the ``$-$'' component,
expands our NLO singular terms
$(\rho)^k I_{k+1}(\sigma)$ in the vicinity of the point
$\sigma = \sigma_{LO} $ and ignores the NLO regular terms (i.e. put  
$exp{(-\overline D_{+}(1)p)}=exp{(-D_{-}(1)p)}=1$ and cancel the terms 
proportionals to $\alpha(Q^2_0)$ into the normalization factors $A_g^\pm$
and $A_q^-$). We think, however, that this expansion is not so correct
because it generates NLO corrections of the order of the LO terms.

\item The negative sign of the NLO correction in  $\sigma$
(see Eq. (\ref{a1}))  
makes excellent the agreement of our result with the parametrization of $F_2$
obtained by De Roeck and De Wolf \cite{DRDW}.
Their result is very similar to our LO form
of $f_q^+$ in Eq.(\ref{8.0}) if one replaces  $s_{LO} \to
s_{LO}^{\delta}$ in the definition (\ref{2.5}) of $s_{LO}$. The
value $\delta = 0.708$ has been obtained in the fit to H1 and ZEUS
data. Due to $\delta <1$, it shows less $Q^2$-dependence
than it is predicted by perturbative QCD at LO. This slower
$Q^2$-dependence may be explained naturally by the negative NLO corrections
to $\sigma$ obtained here.

\item
The behaviour of eqs. (\ref{9.10})-(\ref{9}) can mimic a power law shape
over a limited region of $x, Q^2$.
 \begin{eqnarray}
f_a(x,Q^2) \sim x^{-\lambda^{eff}_a(x,Q^2)}
 ~\mbox{ and }~
F_2(x,Q^2) \sim x^{-\lambda^{eff}_{F2}(x,Q^2)}
\nonumber    \end{eqnarray}
The quark and gluon effective slopes
 $\lambda^{eff}_a = -\frac{d}{d \ln z} 
\ln f_a(z,Q^2)$ are reduced by the NLO terms that leads to the decreasing
of the gluon distribution at small $x$. For the quark case
it is not the case, because the normalization factor $A_q^+$ of the ``$+$'' 
component produces an additional contribution undampening as $\sim (lnz)^{-1}$.

\item
The gluon effective slope $\lambda^{eff}_g$ is larger than the quark slope
$\lambda^{eff}_q$, which is in excellent agreement with a recent MRS and 
GRV analyses \cite{MRS,GRV}. \\
Indeed, because $d/dlnx = d/dlnz$,
the effective slopes
have the form,

 \begin{eqnarray}
\lambda^{eff}_g(z,Q^2) &=& \frac{f^+_g(z,Q^2)}{f_g(z,Q^2)} \cdot
\rho \cdot \frac{I_1(\sigma)}{I_0(\sigma)}
\nonumber
\label{10.1}
\\
\lambda^{eff}_q(z,Q^2) &=& \frac{f^+_q(z,Q^2)}{f_q(z,Q^2)} \cdot
\rho \cdot \frac{ I_2(\sigma) (1- \overline d^q_{+-}(1) \alpha(Q^2))
 + 20 \alpha(Q^2) I_1(\sigma)/\rho}{ I_1(\sigma) 
(1- \overline d^q_{+-}(1) \alpha(Q^2))
 + 20 \alpha(Q^2) I_0(\sigma)/\rho}
\\
\lambda^{eff}_{F2}(z,Q^2) &=& \frac{\lambda^{eff}_q(z,Q^2) \cdot
f^+_q(z,Q^2) + (2f)/3\alpha(Q^2)\cdot \lambda^{eff}_g(z,Q^2) \cdot
f^+_g(z,Q^2)}{f_q(z,Q^2) + (2f)/3\alpha(Q^2)\cdot f_g(z,Q^2)}
\nonumber
\end{eqnarray}

The effective slopes $\lambda^{eff}_a $ and 
$\lambda^{eff}_{F2}$ depend on the magnitudes $A_a$ of the initial PD
and also on the chosen input values of $Q^2_0$ and $\Lambda $.
At quite 
large values of $Q^2$, where the ``$-$'' component is not relevant,
the dependence on the magnitudes of the initial PD disappear, having
in this case for the asymptotic values:

 \begin{eqnarray}
\lambda^{eff,as}_g(z,Q^2) &=& 
\rho \frac{I_1(\sigma)}{I_0(\sigma)} \approx \rho - 
\frac{1}{4\ln{(1/z)}} 
\nonumber \\
\lambda^{eff,as}_q(z,Q^2) &=& 
\rho \cdot \frac{ I_2(\sigma) (1- \overline d^q_{+-}(1) \alpha(Q^2))
 + 20 \alpha(Q^2) I_1(\sigma)/\rho}{ I_1(\sigma) 
(1- \overline d^q_{+-}(1) \alpha(Q^2))
 + 20 \alpha(Q^2) I_0(\sigma)/\rho}
 \nonumber \\
&\approx & \biggl( \rho - \frac{3}{4\ln{(1/z)}} \biggr) \biggl(1- 
\frac{10\alpha(Q^2)}{(\hat d_+ s + \hat D_+ p)} \biggr)
\label{11.1} \\
\lambda^{eff,as}_{F2}(z,Q^2) 
&=& \lambda^{eff,as}_q(z,Q^2) 
\frac{ 1 + 6 \alpha(Q^2)/\lambda^{eff,as}_q(z,Q^2)}{ 1 + 
6 \alpha(Q^2)/\lambda^{eff,as}_g(z,Q^2)} + ~O(\alpha^2(Q^2)) 
\nonumber \\
&\approx & 
\lambda^{eff,as}_q(z,Q^2) + \frac{3 \alpha(Q^2)}{\ln(1/z)},
\nonumber
\end{eqnarray}
where
simbol $\approx $ marks approximations obtained by expansions of modified
Bessel functions $I_n(\sigma)$. These aprroximations should be correct only 
at very large $\sigma $ values (i.e. at very large $Q^2$ and/or very small 
$x$).

We would like to note that at LO, where $\rho  = \rho_{LO} $, the
slope $\lambda^{eff,as}_{F2}(z,Q^2)=
\lambda^{eff,as}_q(z,Q^2)$ 
coincides at very large $\sigma $
with one obtained in \cite{Navelet} (see also \cite{CoDeRo})
in the case of flat input. At the NLO approximation the slope 
$\lambda^{eff,as}_{F2}(z,Q^2)$ lies between quark and gluon ones but
closely to quark slope $\lambda^{eff,as}_{q}(z,Q^2)$ (see also Fig. 3).

\item
Both slopes $\lambda^{eff}_a$ decrease with decreasing $z$. 
A $z$ dependence of the slope should not appear
for a PD with a Regge type
asymptotic ($x^{-\lambda}$) and precise measurement of the slope 
$\lambda^{eff}_a$ may lead to the possibility to verify the type of small
$x$ asymptotics of parton distributions.
\end{itemize}

\section{Results of the fits} \indent

With the help of the results obtained in the previous section we have
analyzed $F_2$ HERA data at small $x$ from the H1 \cite{H1}
and ZEUS \cite{ZEUS} collaborations separately.
Initially our solution of the DGLAP equations depends on five
parameters, i.e. $Q_0^2$, $x_0$, $A_q$, $A_g$ and $\Lambda_{\ms}(n_f=4)$.
In order to keep the analysis as simple as possible
we have fixed $\Lambda_{\ms}(n_f=4) = 250$ MeV which
is a reasonable value extracted from the traditional (higher $x$)
experiments and
that has also been used by others \cite{Munich1}.
The initial scale of the PD was also fixed
into the fits to $Q^2_0$ = 1 $GeV^2$, although later it was released
to study the sensitivity of the fit to the variation of this parameter.
The analyzed data region was restricted to $x<0.01$ to remain within the
kinematical range where our results are
accurate. Finally, the number of active flavors was fixed to $f$=4. 

Tab. 2 contains the results of the fits to H1 data using
Eqs. (\ref{8}) at LO and (\ref{9}) at NLO. The errors used in the calculation
of $\chi^2 $ are statistical and systematic added in quadrature.

Fig. 1 shows $F_2$ calculated from the fit
with Q$^2$ $>$ 1 GeV$^2$ given in table 1 in comparison with H1 data.
Only the lower $Q^2$ bins are shown.
One can observe that the NLO result lies closer to the data
than the LO curve.
The lack of agreement between data and lines observed
at the lowest $x$ and $Q^2$ bins suggests
that the flat behavior should occur at $Q^2$ lower
than 1 GeV$^2$.

In order to study this point we have done the
analysis considering $Q_0^2$ as a free parameter. In exchange, we have
fixed $x_0$ to 1. 
In this case, our formulas must be used with
caution because during the fit $Q_0^2$ could eventually reach a very small
value such that
the argument of the square root in equation (26) becomes
negative when the NLO term proportional to $p$ becomes larger
than the LO term proportional to $s$ . The kinematic region where this
problem happens depends on $Q_0^2$ ($Q_0^2$ $<$ $Q^2$ $<$ $Q_1^2$), where
the upper limit $Q_1^2$ also depends on $Q_0^2$.  
For example for the sequence $Q_0^2 = 0.5/0.4/0.3/0.2$ GeV$^2$
$Q_1^2 = 0.7/1/2.3/180$ GeV$^2$.
Thus, in the fits one should be out of this region
where more terms in the perturbative expansion (NNLO) are needed.

Comparing the results of the fits in table 3 with those in table 2
one can notice a significant reduction in the value of
$A_g$, $Q_0^2$ and the $\chi^2$.
In Fig. 1 the better agreement with the experiment of the
NLO curve is apparent at the lowest kinematical bins.

Tab. 3 also contains the results of the combined fit to 3 different
data set of ZEUS \cite{ZEUS}. The LO and NLO results are compared
with data in Fig. 2. The main conclusion is that
the parameters extracted from fits to H1 and ZEUS are very similar.

The observed decreasing of the gluon density at small $Q^2$
agrees with the 
traditional assumption (see \cite{Reya})
that the main contribution at small
$Q^2$ comes from valence quarks. Thus, this non-zero 
contribution of singlet quark distribution at small $Q^2$ seems to mimic the
valence quark distribution.
Indeed, in our approach the nonsinglet quark distribution is omitted and this
approximation maybe is not so correct at very small $Q^2$.
As it was shown
in the analysis of Ref. \cite{EMR} based on BFKL dynamics, the nonsinglet quark 
distributions has the following small $x$ asymptotic behavior:
 \begin{eqnarray}
xf_{NS}(x, Q^2) \sim x^{1-a_{NS}} ~~~ \mbox { with }~~~
a_{NS} \sim \sqrt{\frac{32}{3} \overline \alpha_{LO} },
 \label{ns1}
 \end{eqnarray}
where
$\overline \alpha_{LO}$ is a $Q^2$-independent coupling constant. The numerical
estimations with the running coupling (see \cite{EMR}) lead to the behaviour 
similar to (\ref{ns1}) with $\overline \alpha_{LO}(Q^2)$ having some effective
energy scale $\Lambda$ proportional $\Lambda_{\ms}$.

At very small $Q^2_0$ (and $x \sim 10^{-2}$) the value of $a_{NS} $ may reach 
the large value $a_{NS} \sim 1$, and the NS quark distributions may start to
contribute together with the gluon and the singlet distributions.
Unfortunately,
it is very difficult to add the NS quark distribution to our analysis based on
DGLAP evolution. Indeed, in agreement with DGLAP dynamics, the non-singlet
quark distribution with $a_{NS} \sim 1$ should be $Q^2$-independent
which seems
contrary to its functional form (\ref{ns1}) with the running coupling 
$\overline \alpha_{LO}(Q^2)$.

Finally with the help of Eq. (\ref{10.1}) we have estimated
the $F_2$ effective slope using the value of the parameters extracted
from NLO fits to data. For H1 data we found
$0.05 < \lambda^{eff}_{F2} < 0.30-0.37$ and for ZEUS   
$0.07-0.09 < \lambda^{eff}_{F2} < 0.31-0.34$. The lower (upper) limits
corresponds to $Q^2=1.5$ GeV$^2$ ($Q^2=400$ GeV$^2$). The dispersion
in some of the limits is due to the $x$ dependence.
Fig. 3 shows that the three types of asymptotical slopes have
similar values.
The NLO values of
$\lambda^{eff,as}_{F2}$ lie between the quark and the gluon ones but
closer to the quark slope $\lambda^{eff,as}_{q}$.
These results are in excellent agreement with those obtained by others 
(see references \cite{H1,MRS,GRV,Navelet} and also the review
\cite{CoDeRo} and references therein).

\section{Conclusions} \indent

We have presented the rules to construct the small $x$ form of parton 
distributions having soft initial conditions at the first two orders of 
perturbation theory. The rules are based on the
exact $n$-space solution and lead to the possibility
of avoiding complicated methods for evaluating 
the inverse Mellin convolution or  special analyses of DGLAP equations
(see \cite{BF2} for a review on this methods).
We have presented here the PD form at LO and NLO approximations of 
perturbation theory, where the corresponding exact solutions in $n$-space
are fully known. Our expressions have quite simple form and reproduce many
properties of parton distributions at small $x$,
that have been known from global fits.

We found the very good agreement between our approach based on QCD at
NLO approximation and HERA data, as it has been observed earlier with
other approaches (see the review \cite{CoDeRo}). Thus, the nonperturbative
contributions as shadoving effects \cite{Levin}, higher twist effects
\cite{Bartels} and others seems to be quite small or seems to canceled
between them and/or with $ln(1/x)$ terms containing by higher orders of
perturbative theory. To clear up the correct contributions of nonperturbative
dynamics and higher orders containing strong $ln(1/x)$ terms, it is necessary
as more precise data and futher effors in developing of theoretical
approaches.\\

It is very useful to evaluate the derivatives of $F_2$ and the parton
distributions with respect to the logarithms of 1/$x$ and $Q^2$
directly from eqs.(\ref{9.10})-(\ref{9}).
We have presented the result for the $x$ logarithmic dependence. 
The calculation of $dF_2/dln(Q^2)$ and $df_a/dln(Q^2)$
is also extremely important  in view of the recent claim
of disagreement between data and predictions in
perturbative QCD \cite{Burd} (see, however, new set GRV analysis \cite{GRV},
where this disagreement decreases essentially).
We are considering to present this work and also predictions for SF $F_L$
in a forthcoming article \cite{all}.

\vspace{1cm}
\hspace{1cm} \Large{} {\bf Acknowledgements}    \vspace{0.5cm}

\normalsize{}

One of the authors (G.P.) was supported in part by Xunta de Galicia
(XUGA-20604A96) and CICYT (AEN96-1673).

\section{Appendix} \indent
\def\theequation{A\arabic{equation}}
\setcounter{equation}0

  For reader convenience  we present here the illustration\footnote{
Contrary to Ref. \cite{method} we use here the variable $z=x/x_0$.}
of the method to 
replace the
convolution of two functions
by a simple product at small$~x$. We limite ourselves to the case of regular
behavior of kernel moments at $n \to 1$.
More detailed analysis can be found in \cite{method} and in \cite{all}.

Let us to consider the set of PD with have the different forms:
\begin{itemize}
\item Regge-like form $f_R(z)=z^{-\delta}\tilde f(z)$, 
\item Logarithmic-like form $f_L(z)=z^{-\delta} ln(1/z) \tilde f(z)$, 
\item Bessel-like form $f_I(z)=z^{-\delta} I_k(2\sqrt{\hat d ln(1/z)})
\tilde f(z)$, 
\end{itemize}
where $\tilde f(z)$ and its derivative $\tilde f'(z)
\equiv d\tilde f(z)/dz$ are smooth at $z=0$ and both are equal to
zero at $z=1$:
  \begin{eqnarray}
\tilde f(1) ~=~ \tilde f'(1) ~=~0 \nonumber
 \end{eqnarray}

  {\bf 1.} Consider the basic integral with even $n > 1$:
  $$J_{\delta,i}(n,z)=z^n \otimes f_i(z) \equiv
  \int_{z}^{1}\frac{{\rm d}y}{y}~y^n~ f_i\biggl(\frac{z}{y}\biggr)~,
~~~~ i=R,L,I$$

{\bf a)} {\it Regge-like case}.
Expanding $\tilde f(z)$ near
$\tilde f(0)$ ,
 we have
 \begin{eqnarray} J_{\delta,R}(n,z) &=&
 z^{-\delta}\int_{z}^{1}{\rm d}y~ y^{n+\delta-1} \left[
 \tilde f(0)+\frac{z}{y}~ \tilde f^{(1)}(0)+ \ldots
  +\frac{1}{k!} \left(\frac{z}{y}\right)^k \tilde f^{(k)}(0)+ \ldots
 \right] \nonumber  \\
&=& z^{-\delta}
 \left[ \frac{1}{n+\delta}~ \tilde f(0)+
 O(z) \right]  \label{A1}  \\
&-& z^n
 \left[ \frac{1}{n+\delta}~ \tilde f(0)+
  \frac{1}{n+\delta-1}~ \tilde f^{(1)}(0)+ \ldots
  +\frac{1}{k!}~ \frac{1}{n+\delta-k}~ \tilde f^{(k)}(0)+ \ldots
 \right] \nonumber
\end{eqnarray}
The second term on the r.h.s. of Eq.(\ref{A1}) can be summed:
$$J_{\delta,R}(n,z)=z^{-\delta}
\left[\frac{1}{n+\delta}~ \tilde f(0)+
O(z) \right]~ + ~
z^n~ \frac{\Gamma(-(n+\delta))\Gamma(1+\nu)}{\Gamma(1+\nu-n-\delta)}~
\tilde f(0)$$

Because our interest here is limited by the nonsingular case ($n \geq 1$),
we can neglect the second term and obtain:

$$J_{\delta, R}(n,z)=z^{-\delta}\frac{1}{n+\delta}~
\tilde f(
z) +
O(z^{1- \delta})$$

{\bf b)} {\it Logarithmic-like case}.
Using the simple relation
$z^{-\delta} \ln(1/z) = (d/d\delta) z^{-\delta}$ we immediately obtain
 \begin{eqnarray}
J_{\delta,L}(n,z) &=&
z^{-\delta} \ln(1/z)
 \Biggl[ \frac{1}{n+\delta} 
\biggl(1- \frac{1}{(n+\delta)\ln(1/x)} \biggr)
~ \tilde f(0)  ~+~  O(z) \Biggr] \nonumber \\
&=& \frac{1}{n+\delta} 
\biggl(1- \frac{1}{(n+\delta)\ln(1/z)} \biggr)
~ f_L(z)  +  O(z^{1-\delta}) \nonumber \\ &=&
\frac{1}{n+\delta} ~ f_L(z)  +  O(1/ln(1/z))
\nonumber
\end{eqnarray}

{\bf c)} {\it Bessel-like case}.
Representing Bessel function in the form
$$
I_k(2\sqrt{\hat d ln(1/z)}) ~=~
\sum^{\infty}_{n=0} \frac{1}{n!}\frac{1}{\Gamma(n+k+1)} 
{\biggl(\hat d \frac{d}{d\delta}\biggr)}^n z^{-\delta}{\biggr|}_{\delta=0} $$
and repeating the above analysis, we have
 \begin{eqnarray}
J_{\delta,I}(n,z) =
\frac{1}{n+\delta} ~ f_I(z)  +  O\Biggl(\sqrt{\frac{\hat d }{ln(1/z)}}\Biggr)
\nonumber
\end{eqnarray}

  {\bf 2.} Consider the integral
$$I_{\delta}(z)=\tilde K(z)\otimes f(z) \equiv
 \int_{z}^{1}\frac{{\rm d}y}{y}~\hat K(y)~ f(\frac{z}{y})$$
and define the moments of the kernel $\tilde K(y)$ in the following form
$$K_n= \int_{0}^{1}{\rm d}y~y^{n-2}\tilde K(y)$$

In analogy with subsection {\bf 1} we have for the Regge-like case:
 \begin{eqnarray} I_{\delta,R}(z) &=&
 z^{-\delta}\int_{z}^{1}{\rm d}y~ y^{\delta-1}~\tilde K(y)~ \left[
 \tilde\varphi(0)+\frac{z}{y}~ \tilde\varphi^{(1)}(0)+ \ldots
  +\frac{1}{k!} \left(\frac{z}{y}\right)^k \tilde\varphi^{(k)}(0)+ \ldots
 \right] \nonumber  \\
&=& z^{-\delta}
 \left[ K_{1+\delta}~ \tilde\varphi(0)+
 O(z) \right]  \nonumber \\
&-& 
 \left[ N_{1+\delta}(x)~ \tilde\varphi(0)+
  N_{ \delta}(z)~ \tilde\varphi^{(1)}(0)+ \ldots
  +\frac{1}{k!}~ N_{1+\delta-k}(z)~ \tilde\varphi^{(k)}(0)+ \ldots
 \right], \nonumber
\end{eqnarray}
where
$$N_{\eta}(z)= \int_{0}^{1}{\rm d}y~y^{\eta-2}\tilde K(zy)$$

The case $K_{1+ \delta} = 1/(n + \delta)$ corresponds to $\tilde
K(y)=y^n$
and has
been already considered in subsection { \bf 1}. In the more general cases
(for example, $K_{1+ \delta} = \Psi(1+ \delta) + \gamma$) we can represent
the "moment" $K_{1 + \delta}$ as 
series of the sort
$ \sum_{m=1} 1/(n + \delta + m)$.

So, for the initial integral at small $x$ we get the simple equation:

$$I_{\delta,R}(z)=z^{-\delta}~  K_{1+\delta}~
\tilde f(
z) +
O(z^{1- \delta}) =  K_{1+\delta}~ f_R(z) + O(z^{1- \delta})$$

Repeating the analysis of the subsections (1b) and (1c), one easily obtains

 \begin{eqnarray}
I_{\delta,L}(n,z) &=& K_{1+\delta}~ f_L(z) + O\biggl(\frac{1}{ln(1/z)}\biggr)
~\mbox{ and }~ \nonumber \\
I_{\delta,I}(n,z) &=& K_{1+\delta}~ f_I(x) 
+  O\Biggl(\sqrt{\frac{\hat d }{ln(1/z)}}\Biggr) \nonumber
\end{eqnarray}

Thus, in the nonsingular case the results do not depend on the shape of the
PD but the accuracy of the method decreases for a $ln(1/z)$-dependent PD.

\newpage

\hspace{1cm}  {\Large{\bf Tables}}    \vspace{0.3cm}

\begin{center}
\begin{tabular}{|l||l|l||l|l||l|l|}         \hline
$f$ & $\hat d_{+}$ & $\hat D_{+}$ &  $\overline d_{+}(1)$
   &  $ \overline D_{+}(1)$ &  $d_{-}(1)$
    & $ D_{-}(1)$   \\ \hline
3 & -4/3 & 1180/81 & 101/81 &  -43.37 & 16/81 
  & 1.974  \\
4 & -36/25 & 91096/5625 & 61/45  & -45.49 & 64/225 
  & 3.108  \\
5 & -36/23 & 84964/4761 & 307/207  & -47.73 & 80/207
  & 4.675  \\ 
6 & -12/7 & 8576/441 & 103/63  & -50.05 & 32/63 
  & 6.864  \\
\hline
\end{tabular}
\end{center} 
{{\bf Table 1.} The values of the parameters used in the calculation of
the parton distributions as a function of the number of flavors}

\vspace{0.5cm}

\begin{center}
\begin{tabular}{|l||l|l|l|l|}         \hline
$Q^2>$ & $A_q$ & $A_g$ & $x_0$ & $\chi^2/n.o.p.$ \\ \hline
 LO (H1) &  &  & &  \\ 
 1   & 1.06$\pm$0.07 & 2.46$\pm$0.19 & 0.11$\pm$0.02 & 114/104 \\ 
 3   & 0.96$\pm$0.08 & 2.53$\pm$0.19 & 0.12$\pm$0.02 &  88/92  \\ 
 5   & 0.83$\pm$0.09 & 2.47$\pm$0.18 & 0.15$\pm$0.02 &  40/83  \\ 
8.5  & 0.80$\pm$0.10 & 2.30$\pm$0.18 & 0.18$\pm$0.03 &  23/67  \\ \hline
NLO (H1) &  &  & &  \\ 
   1 & 0.97$\pm$0.08 & 1.30$\pm$0.11 & 0.20$\pm$0.03 & 63/104 \\ 
   3 & 0.91$\pm$0.10 & 1.31$\pm$0.11 & 0.22$\pm$0.03 & 50/92  \\ 
   5 & 0.81$\pm$0.10 & 1.28$\pm$0.11 & 0.26$\pm$0.04 & 27/83  \\ 
 8.5 & 0.84$\pm$0.11 & 1.22$\pm$0.11 & 0.28$\pm$0.05 & 21/67  \\ \hline 
\end{tabular}
\end{center}
{{\bf Table 2.} The result of the LO and NLO fits to H1 (1994) data
for different low $Q^2$ cuts. In the fits $Q_0^2$ is fixed to 1 GeV$^2$}

\vspace{0.5cm}

\begin{center}
\begin{tabular}{|l||l|l|l|l|}         \hline
Aprox. & $A_q$ & $A_g$ & $Q_0^2$ & $\chi^2/n.o.p.$ \\ \hline
LO (H1)  & 1.10$\pm$0.08 & 0.35$\pm$0.06 & 0.55$\pm$0.02& 60/104  \\ 
NLO (H1) & 0.83$\pm$0.09 & 0.21$\pm$0.07 & 0.55$\pm$0.03& 45/104  \\ 
\hline
LO (ZEUS)  & 1.13$\pm$0.07 & 0.28$\pm$0.05 & 0.55$\pm$0.02& 174/126 \\ 
NLO (ZEUS) & 0.85$\pm$0.08 & 0.18$\pm$0.05 & 0.56$\pm$0.02& 143/126 \\ 
\hline
\end{tabular}
\end{center}
{{\bf Table 3.} The results of the fits to H1 and ZEUS (1994) data
at LO and NLO with $Q_0^2$ free.}

\newpage

\vspace{0.5cm}

\hspace{1cm} {\Large{\bf Figure captions}}    \vspace{0.5cm}

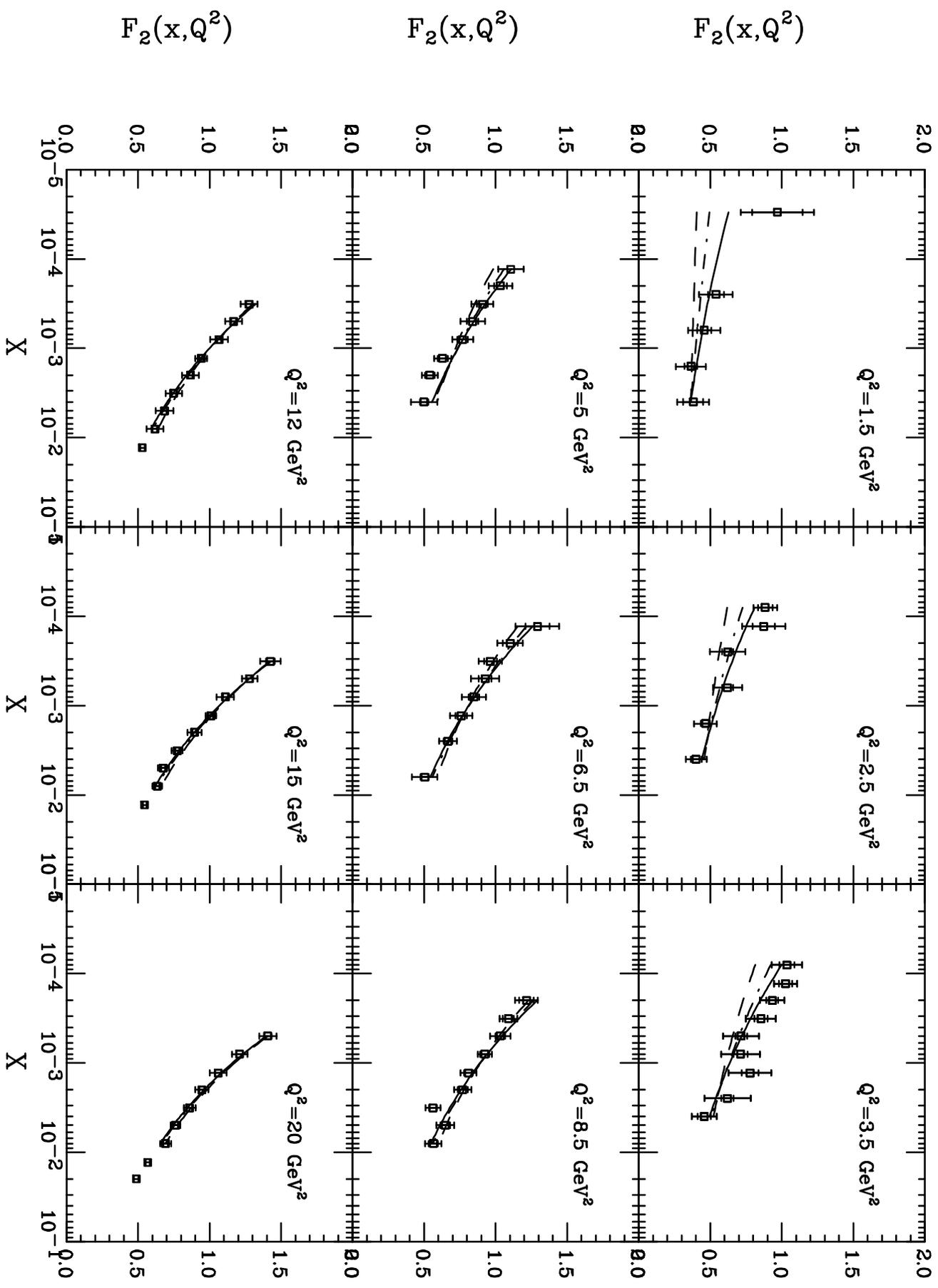
{ \bf Figure 1.} The structure function $F_2$ as a function of $x$
for different $Q^2$ bins. The experimental points
are from H1 \cite{H1}. The inner error bars are statistic while the outer
bars represents statistic and systematic errors added in quadrature.
The dashed and dot-dashed curves are obtained from fits at LO
and NLO respectively with fixed $Q_0^2=1$ GeV$^2$ (see table 2).
The solid line is from the fit at NLO
giving $Q_0^2=0.55$ GeV$^2$ (see table 3).

\vspace{0.5cm}

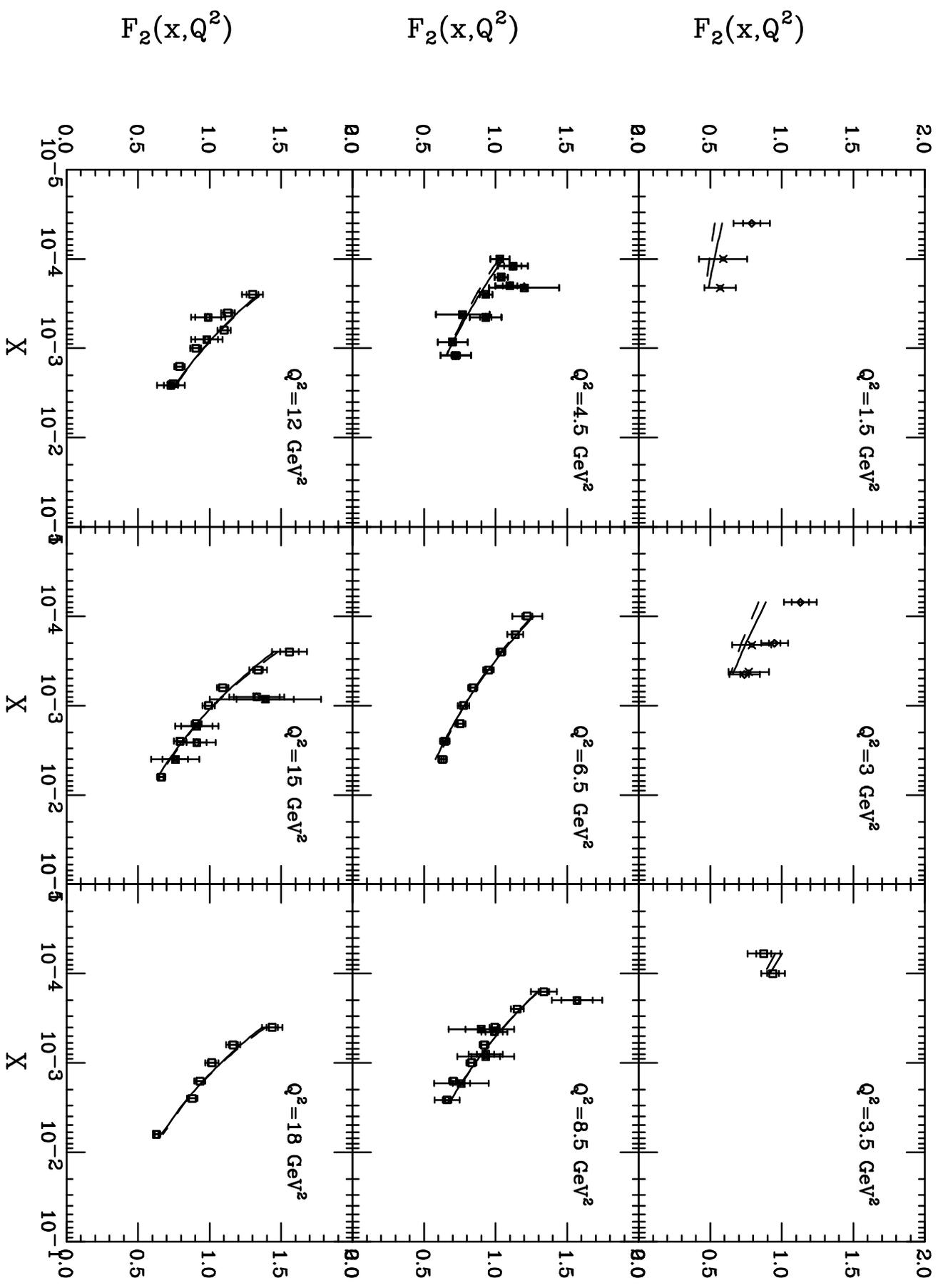
{ \bf Figure 2.} The structure function $F_2$. Experimental points
are from ZEUS (squares from Ref. \cite{ZEUS} and diamonds and crosses
from two different types of measurements reported in Ref. \cite{ZEUSB}).
The error bars are displayed as in Fig. 1.
Solid (dashed) lines are calculated with the parameters given in table 1
from fits at NLO (LO).

\vspace{0.5cm}

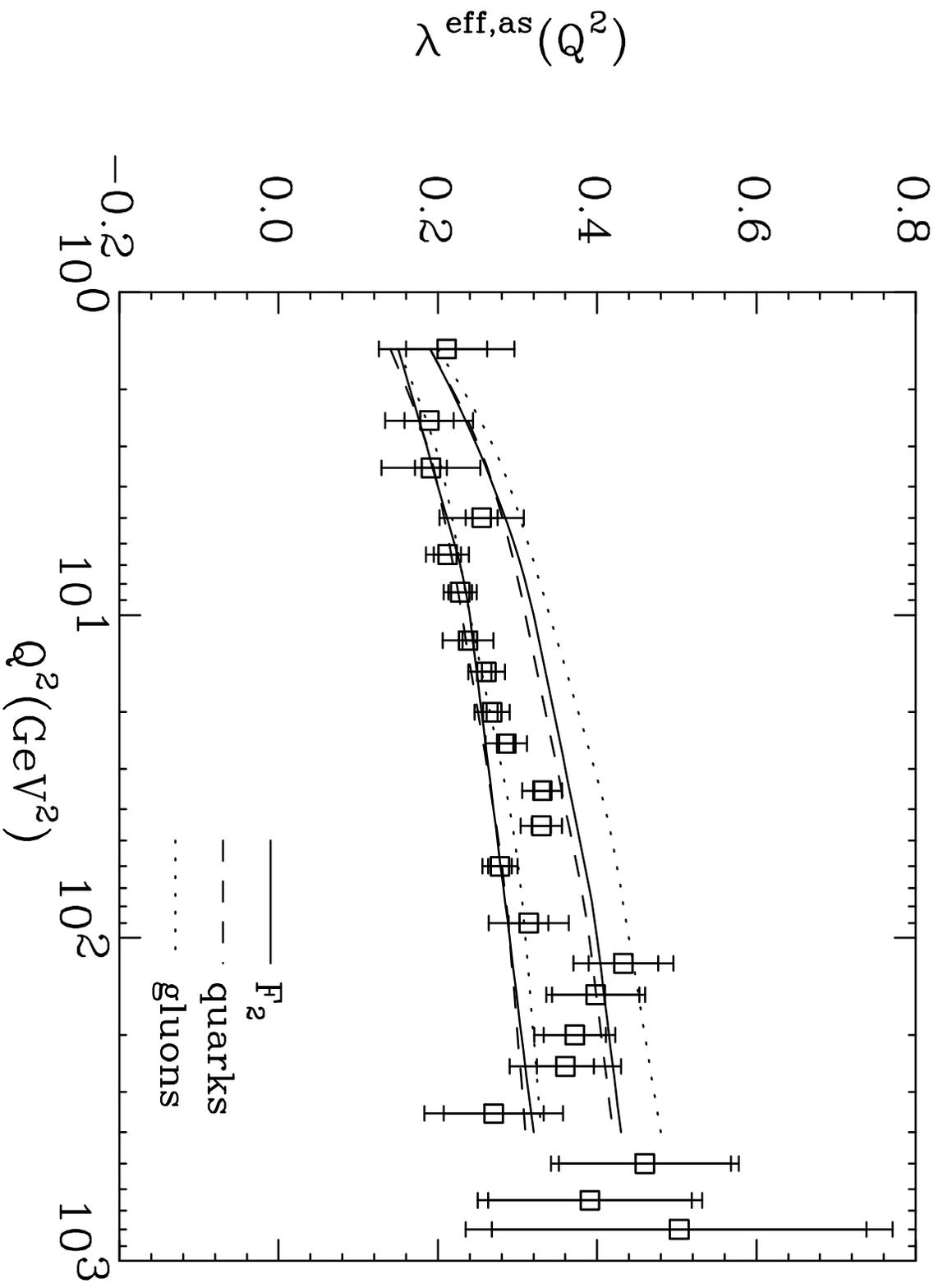
{ \bf Figure 3.} The asymptotical values of effective slopes
$\lambda^{eff,as}_{q}$,
$\lambda^{eff,as}_{g}$ and $\lambda^{eff,as}_{F2}$ calculated at NLO with
the parameters from a NLO fit to H1 with $x_0=1$ (see Tab. (3)).
Lower curves correspond
to $x = 3\times10^{-5}$ while the upper ones are for $x = 10^{-2}$.   
The experimental points are from H1 \cite{H1}.
The error bars are displayed as in Fig. 1.

\end{document}